\documentclass[12pt,a4paper,english]{article}
\usepackage{babel}
\usepackage{indentfirst}
\usepackage{a4wide}

\def\cG{{\cal G}}
\def\cK{{\cal K}}
\def\cH{{\cal H}}
\def\cM{{\cal M}}
\def\cA{{\cal A}}
\def\cB{{\cal B}}
\def\cL{{\cal L}}
\def\cC{{\cal C}}
\def\beq{\begin{equation}}
\def\eeq{\end{equation}}
\def\bea{\begin{eqnarray}}
\def\eea{\end{eqnarray}}

\def\thebiblio#1{
\begin{center}\bf \large References
\end{center}
\list
{[\arabic{enumi}]}{\settowidth\labelwidth{#1.}\leftmargin\labelwidth
 \advance\leftmargin\labelsep
 \usecounter{enumi}}
 \def\newblock{\hskip .11em plus .33em minus -.07em}
 \sloppy
 \sfcode`\.=1000\relax}

\title{{\bf \Large Geometrical formalism in gauge theories}}
\author{Yuri A. Kubyshin$^{1,2}$\footnote{Talk given at the 
CEDYA-03 Congress (Tarragona, September 15-19, 2003)}}
\date{}

\begin{document}

\maketitle

\begin{abstract}
We review the basic elements of the geometrical formalism 
for description of gauge fields and the theory of 
invariant connections, and their applications to 
the coset space dimensional reduction of Yang-Mills 
theories. We also discuss the problem of classification 
of principal fibre bundles, which is important 
for the quantization of gauge theories. Some results 
for bundles over two-dimensional spaces are presented.   
\end{abstract}

\thispagestyle{empty}

\subsection*{Introduction}
\label{Introd}

Nowadays it is a well established fact that the four known 
fundamental interactions of nature - the electromagnetic, weak, 
strong and gravitational - are gauge type interactions 
(see, for example, \cite{FaSl}). This explains the importance of 
gauge field theories which have been the object of intensive 
physical and mathematical studies in the last decades. 

In the present paper we review some results from 
differential geometry and algebraic topology which are 
important for gauge theories. In particular, we discuss  
invariant connections in principle fibre bundles. 
This class of connections includes some monolope 
and instanton solutions 
and is used for description of the gauge theories iimensional base spaces. 
As we will show, the problem of classificatimportant for the functional integral formulation 
of quantum gauge theories.  

The form of the Lagrangian describing quarks and leptons 
is postulated to be locally gauge invariant, i.e. 
invariant with respect to local group transformations 
induced by some group 
$G$ called the gauge group. In models of particle physics $G$ 
is usually assumed to be a compact Lie group. 
Let us denote its Lie algebra by $\cG$. The adjoint actions  
of the group $G$ on itself and on its Lie algebra are 
defined in the standard way: 
for every element $g \in G$ we have
\bea
h & \rightarrow & g h g^{-1} \; \; \; \mbox{for any } \; h \in G, 
\label{ad-G} \\
X & \rightarrow & ad (g) X \equiv g X g^{-1} \; \; \; 
\mbox{for any } \; X \in \cG.      \label{ad-cG} 
\eea

Schematically a theory of gauge interaction is constructed as follows. 
Let $M$ be a Riemannian manifold of dimension $D$ describing the 
space-time and let $\gamma$ be a metric on it. In the 
local coordinates $\{x^{\mu}\}$ it is characterized by the 
metric tensor $\gamma_{\mu \nu}$, where $\mu = 0,1,2, \ldots ,D-1$.  
Suppose that $\psi (x)$ is a matter field, i.e. a field 
describing a lepton or a quark, and its dynamics is described by 
a Lagrangian $\cL_{m} (\psi, \partial_{\mu} \psi)$ 
including only the field function and 
its first derivatives. Depending on its physical characteristics 
(electric charge, leptonic charge, baryonic charge, etc.) the field  
transforms according to 
certain representation $\rho$: for $g \in G$ 
\beq
  \psi (x) \rightarrow \rho (g) \psi (x).    \label{psi-tr-global}    
\eeq
The Lagrangian $\cL_{m} (\psi, \partial_{\mu} \psi)$ is built 
to be invariant under such global transformations. 
This invariance corresponds to the 
conservation of charges associated to the gauge symmetry. 

However, as we already said above, the theory is postulated 
to be invariant under {\it local} gauge transformations, 
i.e. under transformations (\ref{psi-tr-global}) with $g$ which is 
a $G$-valued function $g(x)$. 
An immediate consequence of this is that in order to realize 
such invariance one has to introduce a vector function $A_{\mu}(x)$
with values in the Lie algebra $\cG$ of 
$G$ defined locally, i.e. for each chart of the manifold $M$. 
It is called the gauge potential or gauge field. This is the 
field mediating the interaction of the matter particles 
$\psi (x)$. Under local gauge transformations $A_{\mu}(x)$
and $\psi (x)$ transform as follows:   
\bea
A_{\mu}(x) & \rightarrow & A_{\mu}'(x) = 
ad (g(x)^{-1}) A_{\mu}(x) + 
  g^{-1}(x) \partial_{\mu} g(x),    \label{A-tr} \\
 \psi (x) & \rightarrow & \psi'(x) = \rho (g(x)) \psi (x). 
\label{psi-tr}
\eea
The Lagrangian of the theory, invariant under such 
transformations, is given by 
\beq
\cL (\psi, \partial_{\mu} \psi, A) = \cL_{m} (\psi, D_{\mu}(A) \psi) + 
\cL_{YM} (A), \label{L-tot}
\eeq
where $\cL_{YM}$ is the Yang-Mills Lagrangian, or pure gauge theory 
Lagrangian, invariant under (\ref{A-tr}), and 
\beq
D_{\mu}(A) = \partial_{\mu} + \rho' (A_{\mu}) \label{cov-der}
\eeq
is the covariant derivative. Here $\rho'$ is the representation 
of the Lie algebra $\cG$ corresponding to $\rho$. It can be checked  
that Lagrangian (\ref{L-tot}) is invariant 
under simultaneous gauge transformations (\ref{A-tr}), (\ref{psi-tr}). 
The matter fields interact through the field $A_{\mu}(x)$ by exchanging 
quanta of this field. Thus, the form of the interaction, given by 
$D_{\mu}(A)$, is determined by the local gauge invariance. 
Configurations $(A_{\mu}, \psi)$ and $(A_{\mu}', \psi')$ related by 
(\ref{A-tr}), (\ref{psi-tr}) are equivalent from the physical point 
of view and describe the same state of the system. 
In practical applications 
in order to fix this ambiguity one chooses a representative at each 
orbit of the gauge group action by imposing a gauge fixing condition. 
An example of such condition is the Lorentz gauge: 
$\partial_{\mu} A^{\mu} (x) = 0$, which is widely used in 
calculations of elementary particle processes. 

In the present article we focus on the gauge part 
of the action $S_{YM}[A]$ determined by the 
Lagrangian $\cL_{YM}$. It is equal to 
\beq
S_{YM}[A] = \int_{M} \sqrt{|\det \gamma_{\mu \nu}|} d^{D}x 
\frac{1}{8e^{2}} \left< F_{\mu \nu} F^{\mu \nu} \right>.    
  \label{YM}
\eeq
Here $e$ is a gauge coupling or charge, a numerical constant 
characterizing the 
intensity of the interaction, $<\cdot,\cdot>$ is the Killing form in 
$\cG$ and the gauge field tensor $F_{\mu \nu}$ is defined as  
\beq
F_{\mu \nu}=\partial_{\mu} A_{\nu} - \partial_{\nu} 
A_{\mu} - [A_{\mu}, A_{\nu}]. \label{Fmn-def}
\eeq
For example, the Maxwell electrodynamics is described 
by action (\ref{YM}) with the abelian gauge group $G=U(1)$, 
$A_{\mu}$ is the electromagnetic potential describing photons.   
In the case of the 
Weinberg-Salam-Glashow model, unifying the electromagnetic and 
weak interactions, the gauge group is non-abelian and is 
equal to $G=SU(2) \times U(1)$, certain combinations of the 
$A_{\mu}$ fields describe the $W^{\pm}$-bosons, $Z$-boson and 
photon. 

The plan of the article is the following. 
First we discuss the geometrical 
formalism for the description of the gauge field. Then we 
introduce the notion of invariant connection and review 
main results on their classification. We also outline   
the application of these results to the dimensional reduction 
of multidimensional gauge theories and present a 
concrete example of invariant connection. Finally, we discuss 
briefly the classification of fibre bundles which 
is relevant for the quantization of gauge theories. 

\subsection*{Geometrical formalism for description of gauge theories} 
\label{geometry}

The formalism sketched in the previous section is the one 
usually used in perturbative calculations of concrete 
physical processes and 
effects in the theory of particle interactions. 
However, some properties of the gauge theories are intimately 
related to the geometrical and topological structure of the 
space-time manifold and of the gauge group and are 
not accessible within the perturbation theory expansion. 
Instanton and monopole solutions, a complicated structure 
of the vacuum states, and Chern-Simons models are only few 
examples of this kind \cite{ChengLi}. 

To study such properties of gauge theories another formalism turns 
out to be more adequate. We describe it in this section. 

Let us introduce first the gauge 1-form $A=A_{\mu}dx^{\mu}$ for  
each chart of the space-time manifold $M$.  
Local gauge group transformations (\ref{A-tr}) are written then 
as 
\[
A \rightarrow A' = ad (g^{-1}) A + g^{-1} dg,     
\]
where $ad$ is the adjoint action of the Lie 
group $G$ on its Lie algebra defined by Eq. (\ref{ad-cG}). 
Yang-Mills action (\ref{YM}) can be written as 
\beq
S_{YM}[A] = \int_{M} \frac{1}{4e^{2}} \left< F \wedge *F \right>,   
\label{YM-form}
\eeq
where the 2-form $F$ is determined by the gauge field tensor 
$F_{\mu \nu}$:  
\beq
F = \frac{1}{2} F_{\mu \nu} dx^{\mu} \wedge dx^{\nu},   \label{F-form}
\eeq
and $*$ denotes the Hodge star operation with respect to a 
given metric $\gamma$ on $M$. 

A coordinate independent (and mathematically more 
elegant) description of gauge field is obtained within the 
geometrical formalism (see, for example, \cite{DNF}, \cite{EGH}, 
\cite{NaSe}, \cite{Tra}). 
The basic ingredients are the principal fibre bundle 
$P=P(M,G)$ with the base being the space-time manifold $M$ and 
the structure group being the gauge group $G$ (see, for example, 
\cite{Isham89}, \cite{KN}). 
Let us denote the (free) action $P \rightarrow P$ of the structure 
group on $P$ by $\Psi_{g}$ ($g \in G$) and the canonical 
projection $P \rightarrow M$ by $\pi$. By definition, the 
principal fibre bundle is locally trivial, i.e. every  
point $x$ of $M$ has an open neighborhood $U \subset M$ 
such that $\pi^{-1}(U)$ is homeomorphic to $U \times G$. 
Let us denote the corresponding diffeomorphism 
$\pi^{-1} (U) \rightarrow U \times G$ by $\chi$. It is given by 
a formula: $\chi (p) = (\pi (p),\varphi (p)) \in U\times G$, 
$p \in \pi^{-1}(U) \subset P$, where $\varphi$ is a mapping of  
$\pi^{-1}(U)$ into $G$ satisfying 
$\varphi (\Psi_{g}p) = \varphi (p) g$ for any $p \in \pi^{-1} 
(U)$ and any $g \in G$. 
The action $\Psi_{g}$ of the structure group 
$G$ on $P$ defines an isomorphism $\sigma$ of the Lie algebra 
$\cG$ of $G$ onto the Lie algebra of vertical vector fields on $P$ 
tangent to the fibre at each $p \in P$. 

Depending on the base space and 
structure group the bundle $P$ admits certain connections. 
Gauge fields are described by the connection 1-forms $w$ of 
the connections in $P$. 
To establish a relation with the formalism in terms of gauge 
potentials or gauge 1-forms, described above, one chooses a 
local section $s$ over each open set $U$ of an open covering 
of the manifold $M$, 
$s: U \rightarrow \pi^{-1} (U) \subset P$. Choosing a family  
of sections for an open covering of $M$ 
corresponds to choosing a gauge condition. 
Then the $\cG$-valued gauge 1-form $A^{(s)}$ on $U$ 
is the pull-back of the connection 1-form $w$ on $P$ with 
respect to the 
section $s$: $A^{(s)} = s^{*} w$.
The 2-form $F$ in Eq. (\ref{YM-form}), (\ref{F-form}) 
is calculated as $F^{(s)}=s^{*} \Omega$, 
where $\Omega = Dw$ is the curvature 2-form of the 
connection form $w$. 

It is instructive to see this relation in more detail. 
For this let us consider local coordinates 
$\{x^{\mu}\}$ on $U$ and the local basis of vector fields 
$\xi_{\mu} = \partial_{\mu} \equiv \partial / \partial x^{\mu}$. 
Suppose now that $a$ is a fixed element of the structure group 
$G$ and define a section $s_{a}$ over $U$ as follows: for any 
$x \in U$ $s_{a}(x) = \chi^{-1} (x,a) \in \pi^{-1}(U)$. 
We denote by $(\bar{\xi}_{\mu})_{\chi^{-1}(x,a)}$ the vector field 
$s_{a}' \xi_{\mu}$ tangent to the submanifold 
$\chi^{-1} (U,a)$ of $P$ and by $(\bar{dx}^{\mu})_{\chi^{-1}(x,g)}$ 
the co-vector (1-form) dual to it: 
$\bar{dx}^{\mu} (\xi_{\nu}) = \delta^{\mu}_{\nu}$.   
It can be checked that 
the connection 1-form $w$, which corresponds to the gauge 
potential $A_{\mu}^{s_{a}}(x)$ on $U$, is equal to 
\beq
w = w_{0} + \sigma (A_{\mu}^{s_{a}}(x)) \bar{dx}^{\mu},  
\label{w-A}
\eeq
where $\sigma$ is the mapping from $\cG$ onto the Lie algebra 
of vector fields on $P$ mentioned above. The canonical part $w_{0}$ 
of the connection form is given by the canonical $\cG$-valued 
left invariant 1-form $\theta$ on group $G$ and 
the mapping $\varphi: \pi^{-1}(U) \rightarrow G$ 
which forms part of the diffeomorphism $\chi$ introduced earlier:  
$w_{0} = \phi^{*} \theta$. When the group $G$ is realized by matrices 
$w_{0}$ is often written as 
\[ 
w_{0} = \varphi (p)^{-1} d \varphi(p). 
\]   
Eq. (\ref{w-A}) gives the explicit (local) expression of the 
connection 1-form in terms of the gauge potential 
$A_{\mu}^{s_{a}}$ in the gauge 
corresponding to the section $s_{a}$. 
In this way $A_{\mu}^{s_{a}}$ 
determines the connection in $P$. For example, 
the horizontal lift $\xi^{*}_{\mu}$ of the vector field 
$\xi_{\mu}$ is equal to
\[
\xi^{*}_{\mu} = \bar{\xi}_{\mu} - \sigma (A^{s_{a}}_{\mu} (x)).
\]   

Suppose now that $s$ is a local section in $P$ over $U$ 
and $g(x)$ is a $G$-valued function on $U$. Then $t(x) = 
\Psi_{g(x)} s(x)$ ($x \in U$) defines another section in $P$. 
By straightforward calculation one can show that the gauge 
potentials defined by these two sections are related by 
gauge transformation (\ref{A-tr}) with the function $g(x)$: 
\[
A_{\mu}^{t}(x) = ad (g(x)^{-1}) A_{\mu}^{s}(x) + 
  g^{-1}(x) \partial_{\mu} g(x).   
\]
Thus, indeed, the choice of a local section 
$U \rightarrow \pi^{-1} (P)$ corresponds to the gauge fixing 
in the "physical" formalism described in the Introduction. 
Changing from one section to another is equivalent to a 
gauge transformation and to changing the gauge fixing condition. 
Non-existence of a global cross section 
of a non-trivial fibre bundle $P(M,G)$ means the impossibility 
to introduce the gauge fixing globally. We are forced to use a local 
description of the gauge theory. Contrary to this, the connection 
1-form $w$ exists globally. It contains all the information 
about the gauge field configuration in the theory. Therefore, 
the fibre bundle $P(M,G)$ and $w$ are the objects which define 
completely the gauge theory within the geometrical approach. 

\subsection*{Invariant connections and dimensional reduction} 
\label{invariant} 

The Standard Model, which describes the electromagnetic, weak and strong 
interactions, has been successfully tested in recent particle experiments.  
However, there is a number of important problems which has to be resolved and 
which call for consideration of more general theoretical schemes and 
ideas beyond the Standard Model. One of such approaches, called 
the Kaluza-Klein approach, is based on the 
hypothesis that the space-time has more than four dimensions with 
extra dimensions being a compact space of small enough size $R$. 
Then the dynamics of elementary particles is defined by a 
fundamental multidimensional theory, but at a larger scale 
the additional dimensions are not seen directly and reveal 
themselves in an indirect way through quantum effects.  
This hypothesis does not contradict the observational 
data if $R < 10^{-17}\;$cm. 
By a certain procedure called dimensional reduction a 
multidimensional theory 
can be interpreted in four-dimensional terms, thus 
giving rise to an effective 
theory on the four-dimensional space-time. 

Examples, interesting both from physical and mathematical points of 
view, are given by gauge theories on manifolds $M=M_{4} \times K/H$, 
where $M_{4}$ is the macroscopic 
four-dimensional space-time (for example, the Minkowski space-time) and
the space of extra dimensions is a compact homogeneous space realized as 
a coset space $K/H$, where $K$ and $H$ are compact Lie groups. 
Within the geometrical approach such gauge theory 
is described by connection forms on the principal 
fibre bundle $\hat{P}=P(M,G)$, 
where, as before, $G$ is the gauge group. 

The group $K$ acts on $K/H$ naturally: 
for $k \in K$ and for any point $y=[k_{1}] \in K/H$,  
understood as a class of the coset space $K/H$, the 
transformation is given by $y \rightarrow ky = [k k_{1}]$. 
We assume that $K$ acts trivially on points of $M_{4}$. 
Therefore, the action of $K$ on $M$, which we denote by $O_{k}$, 
can be written as $O_{k} (x,y) = (x, ky)$ for 
$(x,y) \in M_{4} \times K/H$. Moreover, we assume that 
this action can be lifted to the 
fibre bundle $\hat{P}$ as a left bundle 
automorphism\footnote{Actually, in many mathematical 
studies and physical applications the existence of a 
group bundle automorphisms $L_{k}$ with properties (\ref{L1}), 
(\ref{L2}) is taken as the initial assumption, so that the problem 
of the lift of $O_{k}$ is not addressed.}. 
Let us denote this action by 
$L_{k}$. It fulfills the following properties: 
\bea
L_{k_{1}k_{2}} & = & L_{k_{1}} \circ L_{k_{2}},  \label{L1} \\
L_{k} \circ \Psi_{g} & = & \Psi_{g} \circ L_{k}. \label{L2}
\eea
It is said that $K$ is a symmetry group of the gauge 
theory. The transformations $L_{k}$ and $O_{k}$ are related by the 
following formula: 
\[
  \pi \circ L_{k} = O_{k} \circ \pi, 
\]
where $\pi$ is the canonical projection in $\hat{P}$. 
If we denote by $o$ the origin of the coset space 
$K/H$, i.e. the class $o=[e]=H$ containing the group unity $e$, 
then $\tilde{M}_{4} = M_{4} \times \{ o \}$ is 
the submanifold formed by the stable points of the subgroup $H$. 

A connection invariant under the action of the symmetry 
group $K$, or a $K$-{\it invariant connection} in the fibre 
bundle $\hat{P}$ is defined as a connection whose connection 
form $\hat{w}$ satisfies the condition 
\beq
           L^{*}_{k} \hat{w} = \hat{w}.      \label{inv-def}    
\eeq
The invariant connections are interesting from the point 
of view of differential geometry. 
As we will see shortly, they are of special 
importance for the problem of dimensional reduction of 
gauge theories. For consistence one should also consider 
metrics $\hat{\gamma}$ on $M$ which are $K$-invariant:
\beq
     O^{*}_{k} \hat{\gamma} = \hat{\gamma}. \label{gamma-inv}
\eeq
In this case we say that the gauge theory is $K$-invariant. 

Let us discuss the symmetry of gauge potentials which 
correspond to the invariant connections. 
Suppose that $U$ is an open set of a covering   
of the space-time manifold $M$, $s: U \rightarrow \hat{P}$ 
is a local section of $\hat{P}$, and $\hat{A} = s^{*} \hat{w}$ 
is the gauge 1-form on $U$. Under a transformation $O_{k}$ 
a point $\hat{x} = (x,y)$ of $M_{4} \times K/H$ transforms as 
$\hat{x} \rightarrow O_{k} \hat{x}$. 
The invariance condition, Eq. (\ref{inv-def}), gives that 
\beq
O^{*}_{k^{-1}} \hat{A} = ad (\rho_{k})^{-1} \hat{A} + 
(\rho_{k})^{-1} d\rho_{k},  
\label{A-inv}
\eeq
where $\rho_{k}$ is a map $U \rightarrow G$ defined by the 
relation 
\[
L_{k} s(\hat{x}) = \Psi_{\rho_{k}(O_{k} \hat{x})} s (O_{k} \hat{x})
\] 
for any $\hat{x}=(x,y) \in U$. Gauge fields satisfying relation 
(\ref{A-inv}) are called {\it symmetric gauge fields}. 
They were first introduced and studied in papers \cite{Wit77}, 
\cite{Schw}. In Refs. \cite{Schw} the relation between the 
symmetric gauge fields and the invariant connections was 
established and a classification of static spherically-symmetric 
fields in the three-dimensional space was given.    
{}From Eq. (\ref{A-inv}) it follows that for symmetric gauge fields 
the action of the group $K$ is equivalent to a gauge transformation. 
Since Yang-Mills Lagrangian (\ref{YM}) or (\ref{YM-form}) 
is gauge invariant, being calculated on symmetric 
gauge fields it does not depend on points 
of $K/H$ and, therefore, is actually a Lagrangian of an 
effective theory 
on $M_{4}$ only. This feature is the basis of the method 
of dimensional reduction of the sector of 
symmetric gauge fields called the coset 
space dimensional reduction (CSDR). In order to construct 
this effective theory one should express 
$K$-symmetric gauge fields on $M$ in terms of fields on 
$M_{4}$ and then calculate the Yang-Mills action of the 
initial theory in terms of these new four-dimensional fields. 
In this article we consider the first part of this construction. 
We will follow Refs. \cite{KMRV}, \cite{VK-book}.  
In differential geometry it is usually referred to as the 
problem of classification of invariant connections. 
The invariant connections in the mathematical context were 
studied by Wang in Ref. \cite{Wang}, main results can be 
found in book \cite{KN}. In the context of physical gauge 
theories they were extensively studied in a number 
of papers \cite{FoMa}, \cite{CSDR} 
(see \cite{KMRV}, \cite{Zoup} for reviews).

We consider the case when $K$ and $H$ are compact groups. 
Then the homogeneous space is reductive, i.e. the Lie algebra 
$\cK$ of $K$ may be decomposed into a vector space direct sum 
of the Lie algebra $\cH$ of $H$ and an $ad (H)$-invariant 
subspace $\cM$, that is  
\bea
& & \cK  =  \cH \oplus \cM,       \label{cK-decomp} \\
& & ad (H) \cM \subset \cM.     \nonumber 
\eea
The second property implies that 
$[\cH, \cM] \subset \cM$. 

Let us denote the portion of $\hat{P}$ over 
$\tilde{M} = M_{4} \times \{ o \} \cong M_{4}$ as $\tilde{P}$, 
$\tilde{P} = \pi^{-1} (\tilde{M}) \subset \hat{P}$. 
Since transformations $O_{h}$, $h \in H$, act trivially on 
$\tilde{M}$, the transformations $L_{h}$ act on $\tilde{P}$ as 
vertical automorphisms. Therefore, for any 
$\tilde{p} \in \tilde{P}$ there exists a map 
$\tau_{\tilde{p}}: H \rightarrow G$ defined by the relation 
\beq
L_{h} \tilde{p} = \Psi_{\tau_{\tilde{p}}} \tilde{p}. \label{tau-def}
\eeq
It can be easily shown that $\tau_{\tilde{p}}$ is a homomorphism 
of $H$. 

Now let us characterize a $K$-invariant form $\hat{w}$ on $\hat{P}$. 
We denote the inclusion map 
$\tilde{P} \rightarrow \hat{P}$ by $i_{e}$. Then 
$\tilde{w} = i_{e}^{*} \hat{w}$ is a connection 1-form on 
$\tilde{P}$ invariant under the action 
of the group $H$:
\beq
   L^{*}_{h} \tilde{w} = \tilde{w}.    \label{H-inv}
\eeq
This form coincides with $\hat{w}$ in points of $\tilde{P}$ 
and on vector fields tangent to this sub-bundle. The values 
of $\hat{w}$ on other vector fields are given by a 
$\cG$-valued 1-form on $K$ defined by 
$\alpha_{\tilde{p}} \equiv i_{\tilde{p}}^{*} \hat{w}$, where 
the map $i_{\tilde{p}}: K \rightarrow \hat{P}$ is 
induced by the $K$-action on $\hat{P}$ as follows: 
$i_{\tilde{p}} (k) = L_{k} \tilde{p}$. It can be shown that since 
$\alpha_{\tilde{p}}$ is $K$-invariant it is equal to 
\[
\alpha_{\tilde{p}} = \Lambda_{\tilde{p}} (\theta), 
\; \; \; \Lambda_{\tilde{p}} : \cK \rightarrow \cG, 
\]
where $\theta$ is the canonical left invariant 1-form of $K$. 
The map $\Lambda : \cK \rightarrow \cG$ is equivariant, i.e. 
it satisfies the property 
\beq
 \Lambda_{\tilde{p}} \circ ad (h) = ad (\tau_{\tilde{p}}) 
\circ \Lambda_{\tilde{p}}.   \label{L-equiv} 
\eeq
Since the homogeneous space $K/H$ is reductive 
it is enough to specify 
the map $\Lambda_{\tilde{p}}$ for each component of  
decomposition (\ref{cK-decomp}). Its restriction to $\cH$ 
coincides with the differential $\tau'_{\tilde{p}}$ of the 
homomorphism $\tau_{\tilde{p}}$ defined in (\ref{tau-def}):  
\[
\left. \Lambda_{\tilde{p}} \right|_{\cH} = \tau'_{\tilde{p}}: 
\cH \rightarrow \cG,  
\]
whereas its restriction to $\cM$ is given by some 
map $\phi_{\tilde{p}} : \cM \rightarrow \cG$ which is 
equivariant:
\bea
\left. \Lambda_{\tilde{p}} \right|_{\cM} & \equiv & 
\phi_{\tilde{p}},    \nonumber  \\
  \phi_{\tilde{p}} \circ ad (h) & = & ad (\tau_{\tilde{p}}) 
\circ \phi_{\tilde{p}}   \label{phi-equiv} 
\eea
(cf. (\ref{L-equiv})). 

We have obtained that the $K$-invariant form $\hat{w}$ is in 
one-to-one correspondence with the $H$-invariant   
connection 1-form $\tilde{w}$ on $\tilde{P}$ and 
the mapping $\phi_{\tilde{p}}$. The form $\tilde{w}$ characterizes  
the form $\hat{w}$ on the space of orbits of the group $G$ in 
$\hat{P}$, whereas the mapping $\phi_{\tilde{p}}$ characterizes 
$\hat{w}$ along the $G$-orbit passing through $\tilde{p}$. 

Since the form $\tilde{w}$ is still $H$-invariant further 
reduction of the fibre bundle can be carried out, namely the 
reduction of the structure group. This can 
be done in the following way. Since the structure group 
acts freely in each fibre, any two points $\tilde{p}$, 
$\tilde{p}'$ of the same fibre are related by a vertical 
transformation $\Psi_{g}$ for some $g \in G$. Because of 
this the homomorphisms at these points are conjugate to 
each other: 
\[
\tau_{\tilde{p}'} (h) = g^{-1} \tau_{\tilde{p}}(h) g 
\]
for all $h \in H$. Let us fix a point $\tilde{p}_{0}$ in 
$\tilde{P}$ and define 
\[
\tau (h) \equiv \tau_{\tilde{p}_{0}} (h) \; \; \; \;  
\mbox{for all} \; \;  h \in H.
\]
Finally, we denote by $P$ all points $p$ of the fibre 
bundle $\tilde{P}$ for which $\tau_{p} = \tau$. 
It can be shown that $P$ is a 
principal fibre bundle with the base $M_{4}$ and the structure 
group $C = C_{G} (\tau (H))$, i.e. $P = P(M_{4}, C)$. Here 
$C_{G} (\tau (H))$ is the centralizer of the subgroup $\tau (H)$ in 
$G$, 
\[
C_{G} (\tau (H))\equiv \left\{ c \in C | c \tau (h) c^{-1} = 
\tau (h) \; \; \mbox{for any} \; \; h \in H \right\}. 
\]
It can be shown that the restriction of the form $\tilde{w}$ 
to $P(M_{4},C)$, $\tilde{w}|_{P} = w$, is a connection 1-form 
on $P$. It takes values in the Lie algebra $\cC$ of the group 
$C$, as it should be.   

Now we summarize the main result on the classification of 
invariant connections. Any $K$-invariant connection form $\hat{w}$ 
on $\hat{P} (M_{4} \times K/H,G)$ is in one-to-one correspondence 
with the pair $(w,\phi_{p})$ in the reduced bundle $P(M,C)$, 
where $w$ is an arbitrary, i.e. free of additional constraints, 
connection 1-form in $P$ and $\phi_{p}$ is 
a linear equivariant mapping $\phi_{p}: \cM \rightarrow \cG$ 
satisfying property (\ref{phi-equiv}). The explicit relation between 
$\hat{w}$ and $(w,\phi_{p})$ has been described above.  

Let us see how this result is written in terms 
of symmetric gauge fields. 
For this we take an open set $U = U_{M_{4}} \times U_{K/H}$ 
of an open covering of $M$ and a local section 
$s: U \rightarrow \hat{P}$ given by the formula 
\[
s(x,y) = L_{s_{1}(y)} s_{2} (x), \; \; x \in M_{4}, 
\; \; y \in K/H, 
\]
where $s_{1}: U_{K/H} \rightarrow K$ and 
$s_{2}: U_{M_{4}} \rightarrow P$. If $s_{1}$ and $s_{2}$ 
satisfy certain natural conditions, then 
the $K$-symmetric gauge field $\hat{A}$, corresponding to the 
$K$-invariant connection form $\hat{w}$ on $\hat{P}$, is equal to  
\[
\hat{A} (x,y) \equiv (s^{*} \hat{w})_{(x,y)} = 
(s_{2}^{*} w)_{x} + \Lambda_{s_{2}(x)} ((s_{1}^{*} \theta)_{y}), 
\]
where, as before, $\theta$ is the canonical left invariant 1-form of 
the group $K$. Using the properties discussed above the $K$-invarinat 
gauge field can be written as 
\beq
\hat{A}(x,y) = A(x) + \tau'(\bar{\theta}_{\cH}) + 
\phi_{x} (\bar{\theta}_{\cM}). \label{A-repr}  
\eeq
Here $A = s^{*}_{2} w$ is the gauge field form on $M_{4}$, 
$\bar{\theta} = s^{*}_{1} \theta$, $\phi_{x} = \phi_{s_{2}(x)}$, and 
$\bar{\theta}_{\cH}$ and $\bar{\theta}_{\cM}$ denote the $\cH$- and 
$\cM$-components of the form $\theta$, respectively, 
in accordance with decomposition (\ref{cK-decomp}). 
Formula (\ref{A-repr}) 
describes a most general $K$-symmetric gauge field on $M$. 
This is basically Wang's result on classification of 
$K$-invariant connections written in terms of forms on 
the base of the fibre bundle $\hat{P}(M,G)$.  
One can also obtain a general formula for the curvature 
2-form $\hat{\Omega} = d\hat{w} + \frac{1}{2}[\hat{w}, \hat{w}]$ of 
the invariant connection or for the gauge 2-form $\hat{F}$ 
of the symmetric gauge field. Thus, 
\bea
\hat{F} & = & s^{*} \hat{\Omega} = 
F + (D \phi) (\bar{\theta}_{\cM}) + 
\frac{1}{2} [\phi (\bar{\theta}_{\cM}), \phi (\bar{\theta}_{\cM})] 
 \nonumber \\
& - & \frac{1}{2} 
\phi \left([\bar{\theta}_{\cM},\bar{\theta}_{\cM}])_{\cM}\right) 
- \frac{1}{2} 
\tau \left( [\bar{\theta}_{\cM},\bar{\theta}_{\cM}]_{\cH} \right),  
 \label{F-repr}
\eea   
where $[\cdot,\cdot]_{\cM}$ and $[\cdot,\cdot]_{\cH}$ are the 
$\cM$- and $\cH$-components of the product $[\cdot,\cdot]$ in 
the Lie algebra $\cK$ respectively, 
$F=dA + [A,A]$ is the gauge 2-form describing the tension 
of the gauge field $A$ on the four-dimensional 
space-time $M_{4}$, and $D\phi$ is the covariant derivative of 
$\phi$: 
\[
(D \phi)_{x} (\bar{\theta}_{\cM}) = d \phi_{x} (\bar{\theta}_{\cM}) 
+ [A_{x}, \phi_{x} (\bar{\theta}_{\cM})] 
\]
(cf. (\ref{cov-der})). 

To obtain a complete parametrization of invariant connections 
or symmetric gauge fields one has to resolve constraint 
(\ref{phi-equiv}) and construct the mapping $\phi_{x}$. 
This constitutes the second, algebraic part of the CSDR, 
or, equivalently, of the classification problem, 
and is not considered here. Its solution can be found in 
Refs. \cite{KMRV}, \cite{VK-book}. Here we only mention 
that the main idea is to interprete  
constraint (\ref{phi-equiv}) as an intertwining condition and 
$\phi_{x}$ as the intertwining operator which intertwines equivalent 
representations of the group $H$ in $\cM$ and $\cG$. Thus, 
to construct such operator one has to decompose the linear 
space $\cM$ into the irreducuble representations 
of $H$ and $\cG$ into irreducible representations of 
$\tau (H)$ and then use Schur's lemma to construct 
the most general intertwining operator. 

Within the gauge theory the result on classification 
of invariant connections can be interpreted as follows. A  
Yang-Mills theory on the multidimensional space-time 
$M=M_{4} \times K/H$ with the gauge group $G$ considered 
on $K$-symmetric gauge field is reduced to a theory on the 
four-dimensional space-time $M_{4}$ with the 
gauge group $C \subset G$ which includes a gauge field $A$ 
and scalar fields generated by the mapping $\phi_{x}$. 
The action of the reduced theory can be calculated from 
Yang-Mills action (\ref{YM-form}) explicitely using 
representations (\ref{A-repr}), (\ref{F-repr}). The 
concrete form of the reduced action, for example 
the scalar multiplets, the form 
of the potential of the scalar fields and 
explicit expressions of the parameters in 
terms of the multidimensional gauge coupling and the size 
of the space of extra dimensions depend on the geometry 
of the homogeneous space $K/H$, the homomorphism 
$\tau: H \rightarrow G$ and some other properties 
of the groups $G$ and $K$. Results of 
detailed studies of these features, as well as 
a number of interesting examples are presented in 
Refs. \cite{Zoup}, \cite{KMRV}. 
In particular, it was proved that if the space of extra dimensions 
is a symmetric homogeneous space $K/H$, then the scalar potential 
of the reduced theory is always a Higgs type potential and 
leads to the spontaneous symmetry breaking \cite{KMV89}, \cite{Kub91}. 
We would like to mention that the invariant connection can also 
be defined and constructed using the formalism of Lie derivatives 
with respect to the Killing vectors of the extra dimensional 
metric (see Ref. \cite{Zoup}). 

One of possible developments of the formalism of invariant connections 
would be its extension to the case of more general space-time, in particular 
to the case of orbifolds. 
Multidimensional models on $M_{4} \times S^{1}/Z_{2}$  
with a part of physical fields localized on branes situated at the fixed 
points of the orbifold $S^{1}/Z_{2}$ are of much interest due to 
their promissing physical properties.  
Another interesting problem to address is the calculation of 
characteristic classes for invariant connections.  

\subsection*{Example of invariant connection} 

As an illustration of the general construction explained in the 
previous section we will consider the invariant connections 
in the Yang-Mills theory with the gauge group $SU(2)$ on the 
two-dimensional sphere $S^{2}$. 
This example is taken from Ref. \cite{AK}. 

The sphere is realized as a coset space $S^{2} = K/H = SU(2)/U(1)$.
First, let us construct the 1-forms 
$\bar{\theta}_{\cal H}$ and $\bar{\theta}_{\cal M}$
which appear in Eq. (\ref{A-repr}). 
We parametrize the sphere by angles 
$\vartheta$ and $\varphi$ ($0 \leq \vartheta \leq
\pi$, $0 \leq \varphi < 2\pi$) and cover it 
with two charts $(U_{1},\Phi_{1})$ and $(U_{2}, \Phi_{2})$ with 
the open sets $U_{1}$ and $U_{2}$
chosen as $U_{1} = S^{2}-\{\mbox{South pole}\}$ and 
$U_{2} = S^{2}-\{\mbox{North pole}\}$,  
\bea
U_{1} & = & \{ 0 \leq \vartheta < \pi, \; 0 \leq \varphi
< 2\pi \}, \label{chart1} \\
U_{2} & = & \{ 0 < \vartheta \leq \pi, \; 0 \leq
\varphi < 2\pi \}   \label{chart2}
\eea
and $\Phi_{1}$, $\Phi_{2}$ given by means of 
corresponding stereographic projections.  

As generators of $K=SU(2)$ we take
$Q_{j}=\sigma_{j}/2i$ ($j=1,2,3$), where $\sigma_{j}$ are the Pauli
matrices. Let the subgroup $H=U(1)$ be generated by
$Q_{3}$. Then the one-dimensional algebra ${\cal H}$ is spanned by
$Q_{3}$, and the vector space ${\cal M}$ in (\ref{cK-decomp})
is spanned by $Q_{1}$ and $Q_{2}$.
Consider now the decomposition of the algebra ${\cal K}$,
which is the complexification of the Lie algebra of the group
$K=SU(2)$. We denote by $e_{\alpha}$ and
$e_{-\alpha}$ the root vectors and by $h_{\alpha}$ the corresponding
Cartan element of this algebra (see, for example, \cite{GoGro}) 
and take
\[
   e_{\pm \alpha}= \sigma_{\pm} = \frac{1}{2}(\sigma_{1}\pm i\sigma_{2}),
   \; \; \; h_{\alpha} = \sigma_{3}
\]
with
\[
    Ad \left( h_{\alpha} \right) (e_{\pm \alpha}) =
    [h_{\alpha}, e_{\pm \alpha}] =
   \pm 2 e_{\pm \alpha},    \label{h-e}
\]
where $Ad$ denotes the adjoint action of the algebra: 
$Ad (X) Y \equiv [X,Y]$.  
Then ${\cal H}= C h_{\alpha}$ and 
${\cal M} = C e_{\alpha} + C e_{-\alpha}$, and
the decomposition of the vector space ${\cal M}$ into
irreducible invariant subspaces of ${\cal H}$
is described by the following decomposition of the
representations:
\beq
    \underline{2} \rightarrow (2) + (-2),  \label{M-decomp}
\eeq
where in the right hand side we indicated the eigenvalues 
of $Ad (h_{\alpha})$, and the space ${\cal M}$ of the reducible 
representation is indicated by its dimension in the left hand side.

We choose the local representatives $k^{(j)}$ $(j=1,2)$ of
points of the neighbourhood $U_{j}$ of the coset space $S^{2} =
SU(2)/U(1)$ as follows
\[
  k^{(1)}(\vartheta, \varphi) = e^{-i \varphi \frac{\sigma_{3}}{2}}
  e^{i \vartheta \frac{\sigma_{2}}{2}}
  e^{i\varphi \frac{\sigma_{3}}{2}},
  \; \; \;
  k^{(2)}(\vartheta, \varphi) = e^{i\varphi \frac{\sigma_{3}}{2}}
  e^{i(\vartheta-\pi) \frac{\sigma_{2}}{2}}
  e^{-i\varphi \frac{\sigma_{3}}{2}}.
\]
The functions $k^{(i)}: U_{i} \rightarrow SU(2)$ can also be viewed as
local sections of the principal fibre bundle
$K = P(K/H,H)$ over the base $K/H=S^{2}$ with the structure group
$H=U(1)$. By straightforward
computation one obtains the forms $\bar{\theta}_{\cal H}$ and
$\bar{\theta}_{\cal M}$:
\bea
    \bar{\theta}^{(i)} & = & \left(k^{(i)}\right)^{-1} d k^{(i)} =
    \bar{\theta}_{\cal H}^{(i)} + \bar{\theta}_{\cal M}^{(i)},
                                           \nonumber \\
    \bar{\theta}_{\cal H}^{(1)} & = & i \frac{\sigma_{3}}{2}
    \left(1 - \cos \vartheta \right) d\varphi,   \label{theta1} \\
    \bar{\theta}_{\cal M}^{(1)} & = &
    i \frac{\sigma_{1}}{2} \left(-\sin \varphi d\vartheta -
    \sin \vartheta \cos \varphi d\varphi \right) +
       i \frac{\sigma_{2}}{2} \left( \cos \varphi d\vartheta -
       \sin \vartheta \sin \varphi  d\varphi \right),
                                  \nonumber \\
    \bar{\theta}_{\cal H}^{(2)} & = & -i \frac{\sigma_{3}}{2}
    \left(1 + \cos \vartheta \right)
    d\varphi,    \label{theta2} \\
    \bar{\theta}_{\cal M}^{(2)} & = &
       i \frac{\sigma_{1}}{2} \left(\sin \varphi d\vartheta -
       \sin \vartheta \cos \varphi d\varphi \right) +
       i \frac{\sigma_{2}}{2} \left( \cos \varphi d\vartheta +
       \sin \vartheta \sin \varphi d\varphi \right).
                                  \nonumber
\eea

To construct the invariant connection on $M=S^{2}$ we apply 
general formula (\ref{A-repr}). Of course, 
since in the example under consideration the subspace $M_{4}$ 
is absent one should put $A = 0$. 
 
First, let us specify the homomorphism 
$\tau: H = U(1) \rightarrow G=SU(2)$. 
Let $E_{\alpha}$, $E_{-\alpha}$ and $H_{\alpha}$ be the root
vectors and the Cartan element of the algebra ${\cal G} = A_{1}$,
which appears as complexification of the Lie algebra of $G=SU(2)$.
We assume that they are given by the same combinations of the
Pauli matrices as the corresponding elements of complexified
Lie algebra of $K=SU(2)$ described above. The group homomorphism
$\tau$ is given by the expression
\[
   \tau \left( e^{i\frac{\sigma_{3}}{2} \alpha_{3}} \right) =
e^{i \kappa \frac{\sigma_{3}}{2} \alpha_{3}} =
 \cos (\kappa \frac{\alpha_{3}}{2})
  + i \sigma_{3} \sin (\kappa \frac{\alpha_{3}}{2}), 
\]
and it is easy to check that this definition is consistent if
$\kappa$ is integer. Therefore the homomorphisms are
labelled by $n \in Z$. Let us denote them by $\tau_{n}$. 
The induced algebra homomorphism is given by
\beq
     \tau'_{n} (h_{\alpha}) = n H_{\alpha}.  \label{SU2-hom}
\eeq
The three-dimensional space ${\cal G}$ of
the adjoint representation of $A_{1}$ decomposes into three
1-dimensional irreducuble invariant subspaces of
$\tau'_{n}({\cal H})$ and the decomposition
is characterized by the following decomposition of representations:
\beq
    \underline{3} \rightarrow (0) + (2n) + (-2n)  \label{G-decomp}
\eeq
(in brackets we indicate the eigenvalues of
$Ad (\tau'_{n} ( h_{\alpha}))$).

Let us now compare decompositions (\ref{M-decomp}) and
(\ref{G-decomp}). For $n \neq \pm 1,0$ there are no equivalent
representations in the decomposition of ${\cal M}$ and ${\cal G}$ and
the intertwining operator $\phi: {\cal M} \rightarrow {\cal G}$
is zero. It also turns out to be zero for $n=0$. In these cases
according to (\ref{A-repr})
\bea
A^{(1)}_{n} & = &  \frac{i}{2} n \sigma_{3} (1 - \cos \vartheta)
  d \varphi,    \nonumber \\
A^{(2)}_{n} & = &  -\frac{i}{2} n \sigma_{3} (1 + \cos \vartheta)
   d \varphi. \label{A-SU2}
\eea

If $n=1$ or $n=-1$ the results are more interesting.
Let us consider the case $n=1$ first. Comparing (\ref{M-decomp})
and (\ref{G-decomp}) we see that there are pairs of representations
with the same eigenvalues and, therefore, the intertwining
operator $\phi$ is non-trivial. It is determined by its action
on the basis elements of ${\cal M}$:
\beq
   \phi (e_{\alpha}) = f_{1} E_{\alpha}, \; \;
    \phi (e_{-\alpha}) = f_{2} E_{-\alpha},  \label{SU2-phi}
\eeq
where $f_{1}$, $f_{2}$ are complex numbers. The fact that the initial
groups and algebras are compact implies a reality condition \cite{KMRV}
which tells that $f_{1} = f_{2}^{*}$. Thus, the operator
$\phi$ and the invariant connection are parametrized by one
complex parameter $f_{1}$ (we will suppress its index from now on).
Using Eqs. (\ref{theta1}), (\ref{theta2}), (\ref{SU2-hom})
and (\ref{SU2-phi}) we obtain from (\ref{A-repr}) that
\beq
A^{(1)}_{1} =  i \left( \begin{array}{cc}
 (1 - \cos \vartheta) d\varphi & f e^{-i\varphi}
 (-id\vartheta - \sin \vartheta d \varphi) \\
f^{*}e^{i\varphi}(id\vartheta - \sin \vartheta d \varphi) &
 -(1 - \cos \vartheta) d\varphi
     \end{array}  \right),    \label{A-SU2-1}
\eeq
\beq
A^{(2)}_{1} = i \left( \begin{array}{cc}
-(1+\cos \vartheta) d\varphi & f e^{i\varphi} (-id\vartheta -
\sin \vartheta d \varphi) \\
f^{*}e^{-i\varphi}(id\vartheta - \sin \vartheta d \varphi) &
(1+\cos \vartheta) d \varphi
     \end{array}  \right).  \label{A-SU2-2}
\eeq
The curvature form $F = dA + \frac{1}{2} [A,A]$ is described by
the unique expression on the whole sphere and is equal to
\beq
    F  = -  i \sigma_{3} \left( |f|^{2} - 1 \right) \sin
    \vartheta  d\vartheta \wedge d\varphi.      \label{F-SU2}
\eeq
Action (\ref{YM-form}) evaluated for such configuration 
on $M=S^{2}$ is equal to
\beq
   {\cal S}_{YM} (f) = \frac{\pi}{2e^{2}} \frac{1}{R^{2}}
    \left(|f|^{2} - 1 \right)^{2},     \label{S-SU2}
\eeq
where $R$ is the radius of the sphere. Due to the $K$-invariance
any extrema of the action found
withing the subspace of invariant connections is also an
extremum in the space of all connections \cite{KMRV}. From Eq.
(\ref{S-SU2}) we see that there are two types of extrema in
the theory: the maximum at $f=0$ and the minima at
$f$ satisfying $|f|=1$. Only one of them, the trivial
extremum, was found in Ref. \cite{RP} as a spontaneous
compactification solution in six-dimensional Kaluza-Klein theory.

A similar situation takes place for $\kappa=-1$. Again there exists
a 1-parameter family of invariant connections parametrized
by a complex parameter,
say $h$, analogous to $f$. The action possesses two extrema: at
$h=0$ and for $|h|=1$.

It turns out that potentials (\ref{A-SU2}), (\ref{A-SU2-1}) and
(\ref{A-SU2-2}) are related to known non-abelian monopole solutions
in this theory. Namely, for $n \neq \pm 1$ and
for $n = \pm 1$ with $f=0$ these expressions coincide with the monopole
solutions with the monopole number $\kappa = n$. In fact the
solution with $\kappa = n > 0$ can be transformed
to the solution with $\kappa = -n $ by a gauge transformation
$A \rightarrow S^{-1} A S$ with the constant matrix $S=-i \sigma_{1}$.
Eqs. (\ref{A-SU2}) and Eqs. (\ref{A-SU2-1}) and (\ref{A-SU2-2})
with $f=0$ describe all the monopoles in the $SU(2)$ gauge theory
\cite{GNO}. As it was shown in \cite{BrNe}, \cite{Coleman}, all of them,
except the trivial configuration with $n=0$, are unstable.
This is in accordance with the topological classification of monopoles 
by elements of $\pi_{1}(G)$ and also agrees with the fact that 
there is only one bundle (up to equivalence) with the base space
$S^{2}$ and the structure group $SU(2)$. The latter will be 
shown in the next section. 

Thus, all the monopoles are described by connections
in the trivial principal fibre bundle $P(S^{2},SU(2))$ and
can be represented by a unique form on the whole
sphere \cite{CHM}. This is indeed the case.
Namely there exist gauge transformations,
different for $U_{1}$ and $U_{2}$ patches, so that the tranformed
potentials coincide. Let us demonstrate this for the case $n=1$.
In fact this property is true for the whole family of the
invariant connections (\ref{A-SU2-1}), (\ref{A-SU2-2}).
The group elements of these gauge transformations of
the potentials on $U_{1}$ and $U_{2}$  are
\[
V_{1} = i \left( \begin{array}{cc}
 \cos \frac{\vartheta}{2} & e^{-i\varphi} \sin \frac{\vartheta}{2}  \\
 e^{i\varphi} \sin \frac{\vartheta}{2}  & - \cos \frac{\vartheta}{2}
     \end{array}  \right),
\]
\[
V_{2} = i \left( \begin{array}{cc}
 e^{i\varphi} \cos \frac{\vartheta}{2}  & \sin \frac{\vartheta}{2}  \\
 \sin \frac{\vartheta}{2}  & - e^{-i\varphi} \cos \frac{\vartheta}{2}
     \end{array}  \right).
\]
By calculating
\[
   A^{(i)'}_{1}= V_{i}^{-1} A^{(i)}_{1} V_{i} - 
   V_{i}^{-1} d V_{i}
\]
for $i=1$ and $i=2$ one can easily check that the
transformed potentials are equal to each other and are given by
\beq
A^{(1)'}_{1}=A^{(2)'}_{1} = \frac{i}{2} \left( \sigma_{+} c_{+} +
 \sigma_{-} c_{-} + \sigma_{3} c_{3} \right),   \label{A-SU2-gen}
\eeq
where
\bea
c_{+} & = & c_{-}^{*} = e^{-i\varphi} \left\{ \left[
- \cos \vartheta + \left( f \cos ^{2} \frac{\vartheta}{2}
- f^{*} \sin^{2} \frac{\vartheta}{2} \right)
 \right] \sin \vartheta d\varphi
                                         \right. \nonumber \\
 & + & \left. i(-1 + f \cos ^{2} \frac{\vartheta}{2} +
     f^{*}\sin^{2} \frac{\vartheta}{2} ) d\vartheta \right\}
                                  \label{A-SU2-gen1} \\
 c_{3} & = & \left( 1 - \frac{f + f^{*}}{2} \right)
 \sin^{2}\vartheta d\varphi -
i\frac{f-f^{*}}{2} \sin \vartheta d\vartheta.   \nonumber
\eea
Note that in general expressions (\ref{A-SU2-1}), (\ref{A-SU2-2}) and
(\ref{A-SU2-gen}) the phase of the complex parameter $f$ can be
rotated by residual gauge transformations which form the group
$U(1)$.

For $f=0$ this formula gives the known expression for the $\kappa=1$
$SU(2)$-monopole \cite{CHM}:
\beq
A^{(1)'}_{1}=A^{(2)'}_{1} = \frac{i}{2} \left( \begin{array}{cc}
(1-\cos 2\vartheta) d\varphi & e^{-i\varphi}
(-2id\vartheta - \sin 2\vartheta d \varphi) \\
e^{i\varphi}(2id\vartheta - \sin 2\vartheta d \varphi) &
-(1-\cos 2\vartheta) d \varphi
     \end{array}  \right).  \label{A-SU2-0}
\eeq
Of course, forms (\ref{A-SU2-gen}) and (\ref{A-SU2-0})
do not have singularities on the whole sphere.
For $f=f^{*}=1$ forms (\ref{A-SU2-gen1}) vanish. This shows
that this configuration, which is also the extremum of the action,
describes the trivial case of the $SU(2)$-monopole with $\kappa=0$.
Note that in the original form potentials
(\ref{A-SU2-1}) and (\ref{A-SU2-2}) do not seem to be trivial.
Of course, one can check that they are pure gauges and correspond
to a flat connection, i.e. a connection with $\Omega = 0$.  
Vanishing of gauge field 2-form (\ref{F-SU2}) for this value 
of $f$ confirms this.

The picture we have obtained is the following. For different
homomorphisms $\tau_{n}: H \rightarrow G$ we have constructed
different invariant connections given by Eqs. (\ref{A-SU2}),
(\ref{A-SU2-1}) and (\ref{A-SU2-2}). For $n \neq \pm 1$
or $n = \pm 1$ with $f=0$ the connection describes
the $SU(2)$-monopole solution with the monopole number $\kappa=n$.
All $SU(2)$-monopoles on $S^{2}$ are reproduced in this way.
As it was said above the solutions with numbers
$\kappa$ and $(-\kappa)$ are gauge equivalent. In addition, there is a
continuous 1-parameter family of invariant connections which passes
through the configurations describing the $SU(2)$-monopoles with
numbers $\kappa=-1$, $\kappa=0$ and $\kappa=1$
in the space of all connections of the theory. Connections
from this family are described by Eqs. (\ref{A-SU2-1}), (\ref{A-SU2-2}),
(\ref{A-SU2-gen}) and (\ref{A-SU2-gen1}). Not all of these
connections are gauge inequivalent. Classes of gauge equivalent
invariant connections are labelled by values of $|f|$. Thus, $|f|=0$
corresponds to the class of the $\kappa=1$ monopole.
The monopole with $\kappa=-1$ can be obtained from it by the gauge
transformation with the constant matrix $S= - i \sigma_{1}$, as
it was explained above, and, hence, belongs to the same gauge class.
Connections with $|f|=1$ form the class describing the monopole with
$\kappa=0$.

\subsection*{Classification of fibre bundles}
\label{bundles}

So far we have been discussing the classical aspects of gauge theories. 
One of the ways to quantize them consists in considering a 
functional integral of the type  
\beq
Z = {\cal N} \int_{\cal A} {\cal D}A e^{-S_{YM}[A]} T(A),   \label{int}
\eeq
where $S_{YM}[A]$ is the classical action given by (\ref{YM}) or 
(\ref{YM-form}), ${\cal N}$ is a normalization factor and $T(A)$ 
is some function or functional of the gauge field $A$. 
For example, $T(A)$ can be a 
product of $A$-fields at different points or the traced holonomy 
for a closed path in $M$. In the first case integral (\ref{int}) 
defines a Green function, whereas in the second case it gives 
the vacuum expectation value of the Wilson loop variable \cite{ChengLi}. 
The measure ${\cal D}A$ is understood in a 
heuristic sense adopted in quantum field theory. The integral is taken over 
the space $\cA$ of connections. In general $\cA$ may consist of a number of 
components (or sectors) $\cA_{\alpha}$ labelled by elements $\alpha$ of 
an index set $\cB$, $\alpha \in \cB$. Then functional integral (\ref{int}) 
is given by a sum over the elements of $\cB$, each term of the 
sum being the functional integral over the subspace of connections 
$\cA_{\alpha}$. 

The set of connected
components of ${\cal A}$ is in 1-1 correspondence with the set of
non-equivalent (i.e. which cannot be mapped one into another
by a bundle isomorphism) principal $G$-bundles $P(M,G)$ over
manifold $M$. Let us denote this set as ${\cal B}_{G}(M)$,
i.e. ${\cal B} \cong {\cal B}_{G}(M)$. The problem of
characterization of this set and classification of such bundles is
considered in a number of books and articles. A method,
which in many cases gives a solution to this problem and which we
closely follow here, is discussed in lectures \cite{AvIsh} and in 
Refs. \cite{AK}, \cite{Kub}.
Relevant results from algebraic topology can be found in
\cite{Bott,Span,Swit}. Here we only outline the construction.
We would like to note that in fact the considerations and the results
in the rest of this section are valid for a more general case,
namely when $M$ is a path-connected CW-complex and not just a 
smooth manifold, and $G$ is a topological group (see definitions, 
for example, in Ref. \cite{Swit})\footnote{The CW-complexes 
and the homotopy 
equivalences considered here are actually pointed 
CW-complexes and homotopy equivalences between pointed 
spaces. For the sake of simplicity of notations we 
do not indicate the base points explicitely. The reader 
interested in complete formulas is referred to Ref. \cite{Kub}.}. 

It turns out that for any topological group $G$ there exists 
a space $BG$, called the classifying space, and a principal 
$G$-bundle $EG=P(BG,G)$, called the universal $G$-bundle, 
such that every principal $G$-bundle $P(M,G)$ is induced from 
$EG$, i.e. $P = f^{*}(EG)$ for some map 
$f: M \rightarrow BG$ \cite{Swit}. Two homotopic maps 
$M \rightarrow BG$ induce equivalent bundles. The total
space $EG$ of the universal bundle is $\infty$-universal, 
i.e. $\pi_{q}(EG)=0$ for all $q \geq 1$. The universal 
bundle is unique up to homotopy equivalence. Brown's theorem 
\cite{Swit} implies that for any $G$ there exists a 
CW-complex $BG$ 
and a principal $G$-bundle $EG=P(BG,G)$ such that for any 
CW-complex $M$ there is an equivalence  
\[
    {\cal B}_{G}(M) \cong [M \ ;\ BG].
\]
By examing a certain exact homotopy sequence it can be shown
that the base $BG$ of the universal bundle satisfies the
property \cite{AvIsh}, \cite{Swit}
\beq
       \pi_{q}(BG) = \pi_{q-1}(G), \; \; \; q \geq 1.   \label{piBG}
\eeq
As an immediate application of this relation we obtain 
the classification 
of principal $G$-bundles over the sphere $S^{n}$ in terms of 
the $(n-1)$th homotopy group of G:
\beq
\cB_{G} (S^{n}) \cong [S^{n};BG] \cong 
\pi_{n} (BG) \cong \pi_{n-1} (G).    \label{B-Sn}
\eeq
In particular, ${\cal B}_{SU(2)} (S^{2}) \cong \pi_{1} (SU(2)) = 0$, 
i.e. there is only one $SU(2)$-bundle with the base $S^{2}$, 
namely the trivial one $P=S^{2} \times SU(2)$. This result 
has already been mentioned in the previous section.    
 
The question is how one can construct such universal bundles
and characterize $[M;BG]$ in terms of objects which can be calculated
in a relatively easy way. It turns out that the Eilenberg - MacLane
spaces play important role for this problem because of their special
homotopic properties. Such spaces are often denoted as $K(\pi,n)$,
where $\pi$ is a group and $n$ is a positive integer, and are
defined as follows:
\bea
 \mbox{i)} & & \; \; K(\pi,n) \; \; \; \mbox{is path connected};
 \nonumber \\
 \mbox{ii)} & & \pi_{q}\left( K(\pi,n) \right) = \left\{
      \begin{array}{ll}
                 \pi, & \mbox{if} \; \;  q=n, \\
                  0 , & \mbox{if} \; \;  q \neq n.
      \end{array} \right.   \nonumber
\eea
If $n \geq 1$ and $\pi$ is Abelian, the space $K(\pi,n)$
exists as a CW-complex and
can be constructed uniquely up to a homotopy equivalence \cite{Span}.
The property, which is crucially important for us, is the following:
\beq
   [M;K(\pi,n)] \cong H^{n}(M,\pi),   \label{MK-Hn}
\eeq
where $H^{n}(M,\pi)$ is the $n$th singular cohomology group with 
coefficients in $\pi$ \cite{Bred}, \cite{Span},
\cite{Swit}. A simple example is $K(Z,1) = S^{1}$.

Often the Eilenberg-Maclane spaces are infinite dimensional.
The example important for us is the space $CP^{\infty}$ which
is a CW-complex. It is understood as the union (direct limit)
of the complex projective
spaces $CP^{n}$ of the sequence $CP^{1} \subset CP^{2} \subset \ldots $
\cite{Span}. Then $\pi_{q} (CP^{\infty}) = \lim _{j \rightarrow \infty}
\pi_{q} (CP^{j})$ and
\[
\pi_{2}(CP^{\infty}) = Z, \; \; \; \pi_{q}(CP^{\infty}) = 0 \; \;
\mbox{for} \; \; q \neq 2.
\]
Thus, $CP^{\infty} = K(Z,2)$.

This space serves for the classification of principal fiber bundles
with $G=U(1)$ \cite{AvIsh}, \cite{Isham89}.
Indeed, consider the Hopf bundle $S^{2n-1} = P(CP^{n},U(1))$, 
where the sphere is realized as
\[
  S^{2n-1} = \left\{ (z_{1}, \ldots ,z_{n}) \in C^{n} | \sum_{i=1}^{n}
  |z_{i}|^{2} = 1 \right\}
\]
and the bundle projection $p: S^{2n-1} \rightarrow CP^{n}$ is given by
\[
  p (z_{1}, \ldots ,z_{n}) = \left( \frac{z_{2}}{z_{1}}, \frac{z_{3}}{z_{1}},
  \ldots, \frac{z_{n}}{z_{1}} \right)
\]
for a neighbourhood with $z_{1} \neq 0$, etc. Since
$\pi_{q}(S^{2n-1}) = 0$ for $q < 2n-1$, the bundle is $(2n-2)$-universal.
Then one takes the direct limits $S^{\infty} = \lim_{\rightarrow} S^{n}$,
$CP^{\infty} = \lim_{\rightarrow} CP^{n}$. The bundle
$S^{\infty} = P(CP^{\infty},U(1))$ is $\infty$-universal, and, therefore, 
$EU(1) = S^{\infty}$ and $BU(1) = CP^{\infty}=K(Z,2)$. 
Thus, for a CW-complex of {\it any} dimension
\beq
{\cal B}_{U(1)}(M) \cong [M; BU(1)] \cong [M; CP^{\infty}] = [M; K(Z,2)]
\cong H^{2}(M;Z).    \label{BG-U1}
\eeq
This result gives the classification of principal $U(1)$-bundles
$P(M,U(1))$. For the case when $M$ is a smooth manifold it
was discussed in \cite{Kost-AsBo}.

Now let us consider the generalization of this construction
for the case of other gauge groups. The idea is that for classification
of bundles with the base $M$ with $\dim M \leq n$ only homotopy
groups of low dimensions are important. Then instead of $BG$
one can use some other space $BG_{n}$ which is related to it and
which may be easier to construct and to study.
To define $BG_{n}$ let us introduce the notion of $p$-equivalence.
Consider two path connected spaces $X$ and $Y$ and a continuous
map $f: X \rightarrow Y$ such that for all $x \in X$ the induced map
$f_{*}: \pi_{q} (X) \rightarrow \pi_{q}(Y)$
is an isomorphism for $0<q<p$
and an epimorphism for $q=p$. Such map $f$ is called $p$-equivalence.
Then, if there exists some space $BG_{n}$ and a map $f : BG \rightarrow
BG_{n}$ which is $(n+1)$-equivalence, it can be proved that
\[
   [M;BG] \cong [M; BG_{n}]
\]
for any CW-complex $M$ with $\dim M \leq n$ \cite{Bott}, \cite{Swit}.

Another important element which is used for the classification of 
principal fibre bundles is the Postnikov decomposition (called also 
the Postnikov factorization). It allows to construct certain 
exact sequences which give the required $(n+1)$-equivalences. 

{}From now on we restrict the discussion to the case of two-dimensional 
spaces. In addition we assume certain properties of $G$, they are specified 
below. We would like to mention that these properties are not too 
restrictive and are usually fulfilled in physical applications. 
Results for $\cB_{G}(M)$ with $M$ of higher dimensions can be 
found in \cite{AvIsh}. 

Since $\dim M=2$ for the classification of principal $G$-bundles 
it suffices to find a 3-equivalence $BG \rightarrow BG_{2}$. Its 
existence is guaranteed by the Postnikov factorization theorem 
\cite{Bred}, \cite{Span}. Then $\cB_{G}(M) \cong [M;BG] \cong [M;BG_{2}]$. 
The term $[M;BG_{2}]$ appears in an extended Puppe sequence 
in the corresponding Postnikov diagram (see details in \cite{Kub}). 
Finally, we arrive at the following result on the classification 
of principal $G$-bundles over two-dimensional spaces. 

\noindent {\bf Theorem 1.} \cite{Kub} 
{\it Let $M$ be a path-connected pointed CW-complex 
of $\dim M =2$. Let $G$ be a group such that $\pi_{0}(G)$ is abelian and 
discrete and acts trivially on the higher homotopy groups $\pi_{n}(G)$ 
for $n \geq 1$. Then for the set ${\cal B}_{G}(M)$ of equivalence classes 
of principal $G$-bundles over $M$ there exists the following 
short exact sequence of pointed sets:  
\beq
0 \rightarrow H^{2}(M;\pi_{1}(G)) \rightarrow {\cal B}_{G}(M)
\rightarrow H^{1}(M;\pi_{0}(G)) \rightarrow 0.  \label{HBH}
\eeq}

Further specification of ${\cal B}_{G}(M)$ requires the knowledge of
additional information about $M$ or $G$. 
Let us consider two particular cases. 

\begin{enumerate}

\item \underline{$\pi_{1}(G)=0$.}
It follows from (\ref{HBH}) that
\beq
   {\cal B}_{G}(M) \cong H^{1}(M;\pi_{0}(G)).  \label{B-H1}
\eeq
This case includes the class of discrete groups. If $G$ is discrete,
$\pi_{0}(G) \cong G$ and the formula above simplifies further:
\beq
   {\cal B}_{G}(M) \cong H^{1}(M;G).  \label{B-H1a}
\eeq

\item \underline{$G$ is path-connected.}
Then $\pi_{0}(G)=0$ and
\beq
   {\cal B}_{G}(M) \cong H^{2}(M;\pi_{1}(G)).  \label{B-H2}
\eeq
\end{enumerate} 
Let us make a few comments. 
Eq.~(\ref{B-H1a}), giving the classification of principal $G$-bundles with
a discrete structure group, is in fact a particular case of a more
general result valid for CW-complexes $M$ of {\it any} dimension, 
see Ref. \cite{Kub} for further details.  
Relation (\ref{B-H2}) for $G=U(1)$ is, of course, is agreement with 
result (\ref{BG-U1}) for the base space of any dimension.
   
The results in cases 1 and 2 above can be obtained, in fact, from 
the following general theorem.

\noindent {\bf Theorem 2.} \cite{Bred} {\it  
Let $X$ be a pointed CW-complex, $Y$ is an 
$(n-1)$-connected pointed space, and 
$H^{q+1}(X;\pi_{q}(Y))=0=H^{q}(X;\pi_{q}(Y))$
for all $q \geq n$. Then there exists a one-to-one correspondence
\beq
  [X;Y] \cong H^{n}(X;\pi_{n}(Y)).    \label{XY-Hn}
\eeq }

Let us apply this theorem to $X=M$ with $\dim M = 2$. In the 
case when $Y = BG$
is 0-connected with $\pi_{2}(BG)=\pi_{1}(G)=0$, we obtain result
(\ref{B-H1}). If $Y=BG$ is 1-connected, i.e.
$\pi_{1}(BG)=\pi_{0}(G)=0$, then (\ref{XY-Hn}) gives (\ref{B-H2}). The
latter case is also a result of the Hopf-Whitney classification
theorem \cite{Whi}.

Relations (\ref{B-H1}) and (\ref{B-H2}) become more concrete if
further properties of the group $G$ are known.
For completeness of the discussion, we present a list of the
first homotopy groups $\pi_{1}(G)$ for some connected Lie groups:

\begin{enumerate}

 \item Simply connected, $G = SU(n)$, $Sp(n)$: $\pi_{1}(G) = 0$.

 \item $G = SO(n)$, $n = 3$ and $n \geq 5$: $\pi_{1}(G) = Z_{2}$.

 \item $G=U(n)$: $\pi_{1}(G) = Z$.

\end{enumerate}
In particular, using Eq. (\ref{B-H2}) one can obtain that 
$\cB_{SU(2)} (S^{2}) \cong H^{2}(S^{2};\pi_{1}(SU(2)))=0$, 
the result already mentioned above.  

If the space $M$ is of certain type, sequence (\ref{HBH}) 
and relations (\ref{B-H1}), (\ref{B-H2}) become more concrete. 
Thus, if $M$ is a two-dimensional differentiable manifold
one can use known formulas for cohomology groups. For example, if
$\pi$ is abelian then (see \cite{Bred})

\vspace{0.2cm}

\noindent 1) for $M$ compact and orientable 
\beq
H^{2}(M;\pi) \cong \pi,      \label{M-comp}  
\eeq

\noindent 2) for $M$ compact and not orientable 
\beq
H^{2}(M;\pi) \cong \pi/2\pi     \label{M-noncomp}
\eeq
\vspace{0.2cm}

\noindent More expressions for cohomology groups of various 
two-dimensional surfaces can be found in Ref. \cite{GH}. 
In particular, for $M = S^{2}$ it is known that its 
first cohomology group $H^{1}(S^{2},\pi)=0$. Then from 
Theorem 1, Eq. (\ref{MK-Hn}) and the definition 
of the Eilenberg-MacLane spaces it follows that 
\[
\cB_{G} (S^{2}) \cong H^{2} (S^{2}; \pi_{1} (G)) 
\cong [S^{2}; K(\pi_{1}(G),2)] \cong 
\pi_{2} (K(\pi_{1}(G),2))=\pi_{1}(G).
\]
This is in accordance with Eq. (\ref{B-Sn}).    

To finish our discussion of the result let us mention a classification
of principal fibre bundles over two-dimensional compact orientable
manifolds obtained by Witten in Ref. \cite{Wit}.
It is known that any connected Lie group $G$ can be obtained 
as a quotient group $G = \tilde{G}/\Gamma$ (see, for example, \cite{BR}, 
\cite{FH}). 
Here $\tilde{G}$ is the unique (up to isomorphism) connected and simply 
connected Lie group, called a universal covering group of $G$. 
$\Gamma$ is a discrete subgroup of the center $Z(\tilde{G})$ of 
$\tilde{G}$. Witten showed that principal $G$-bundles over $M$ 
are classified by elements of $\Gamma$,
i.e. ${\cal B}_{G}(M) \cong \Gamma$. This agrees with Eq.~(\ref{B-H2}). 
Indeed, taking into account that 
$\pi_{1}(G) \cong \pi_{0}(\Gamma) \cong \Gamma$ and using 
Eq. (\ref{M-comp}), from (\ref{B-H2}) we get 
\[
{\cal B}_{G}(M) \cong H^{2}(M;\pi_{1}(G)) \cong H^{2}(M;\Gamma) \cong 
\Gamma.
\]

\subsection*{Acknowledgements}
This work was supported by the Russian Fund for Basic Research 
(grant 02-02-16444) and the Program "Universities of Russia" 
(grant UR.02.03.002).

\begin{thebiblio}{03}

\bibitem{AK} J.M. Aroca, Yu. Kubyshin, 
"Calculation of Wilson Loops in Two-Dimensional 
Yang-Mills Theories", {\em Annals of Physics}, 
{\bf 283}, (2000), 11-56.

\bibitem{AvIsh} S.J. Avis, C.J. Isham, in {\em Recent Developments in Gravitation. 
Carg\`ese 1978} (M. L\'evy, S. Deser, Eds.), Plenum Press, New York, 1966. 

\bibitem{BR}
A.O. Barut, R. Raczka, {\em Theory of Group Representations and 
Applications}, PWN - Polish Scientific Publishers, Warsaw, 1977. 

\bibitem{Bott}
R. Bott, L.W. Tu, {\em Differential Forms in Algebraic Topology}, 
Springer-Verlag, New York, 1982.

\bibitem{BrNe}
R. Brandt, F. Neri, {\em Nucl. Phys.}, {\bf B161}, (1979), 253.

\bibitem{Bred}
G.E. Bredon, {\em Topology and Geometry}, Springer-Verlag, New York, 1993.

\bibitem{CHM}
Chan Hong-Mo and Tsou Sheung Tsun, {\em Some elementary gauge theory
concepts}, World Scientific, Singapore, 1993.

\bibitem{ChengLi} T.P. Cheng, L.-F. Li, {\em Gauge Theory of Elementary 
Particle Physics}, Oxford University Press, 1984.   

\bibitem{Coleman}
S. Coleman, "The magnetic monopole fifty years later",
in: Proceed. of the Intern. School of Subnuclear Physics, Erice, 1981 
(A. Zichichi, Ed.), Plenum Press, 1983.

\bibitem{DNF} 
B.A. Dubrovin, A.T. Fomenko, S.P. Novikov, {\em Modern Geometry - 
Methods and Applications}, Parts II and III, Springer-Verlag, New York, 1985. 

\bibitem{EGH} 
T. Eguchi, P.B. Gilkey, A.J. Hanson, {\em Physics Reports}, {\bf 66}, (1980), 213. 

\bibitem{FaSl} L.D. Faddeev, A.A. Slavnov, {\em Gauge Fields. 
Introduction to Quantum Theory}, Benjamin-Cummings, Reading 
(Massachusetts), 1980. 

\bibitem{FoMa} P. Forgacs, N.S. Manton, {\em Comm. Math. Phys.}, 
{\bf 56}, (1980), 15. 

\bibitem{FH}
W. Fulton, J. Harris, {\em Representation Theory. A First Course}, 
Springer-Verlag, New York, 1991. 

\bibitem{GNO}
P. Goddard, J. Nuyts and D. Olive, {\em Nucl. Phys.}, {\bf B125}, 
(1977), 1.

\bibitem{GoGro} M. Goto, F.D. Grosshans, {\em Semisimple Lie Algebras}, 
Marcel Dekker, New York, 1978.  

\bibitem{GH}
M.J. Greenberg, J.R. Harper, {\em Algebraic Topology}, 
The Benjamin/Cummings Publishing Co., Inc., Menlo Park, 1981.   

\bibitem{CSDR} 
J. Harnard, S. Shnider, L. Vinet, {\em J. Math. Phys.}, 
{\bf 21}, (1980), 2719-2724. \\
A. Jadczyk, K. Pilch, {\em Lett. Math. Phys.}, 
{\bf 8}, (1984), 97-104. \\
I.P. Volobuev, G. Rudolph, {\em Theor. Math. Phys.}, {\bf 62}, 
(1985), 388-399.  
 
\bibitem{Isham89}
C.I. Isham, {\em Modern Differential Geometry for Physicists}, World Scientific, 
Singapore, 1989. 

\bibitem{Zoup} D. Kapetanakis, G. Zoupanos, {\em Physics Reports}, {\bf C219}, 
(1992), 1.

\bibitem{KN} S. Kobayashi, K. Nomizu, {\em Foundations of Differential Geometry}, 
Interscience, New York, 1963. 

\bibitem{Kost-AsBo}
B. Kostant, in: {\em Lectures in Modern Analysis and Applications III} 
(R.M. Dudley et al., Eds.), Lecture Notes in Mathematics, {\bf 170}, 
Springer, Berlin, 1970, p. 87. \\
M. Asorey, L.J. Boya, {\em J. Math. Phys.}, {\bf 20}, (1979), 2327.

\bibitem{KMV89} 
Yu.A. Kubyshin, J.M. Mour\~ao, I.P. Volobujev, 
"Mul\-ti\-di\-men\-sio\-nal Ein\-stein-Yang-Mills theories: 
dimensional reduction, spontaneous compactification and all 
that", {\em Nucl. Phys.}, {\bf B322}, (1989), 531-554.

\bibitem{KMRV} Yu.A. Kubyshin, J.M. Mour\~ao, G. Rudolph, 
I. Volobujev, {\em Dimensional 
Reduction of Gauge Theories, Spontaneous Compactification and Model Building}, Springer-Verlag, Berlin, 1989. 

\bibitem{Kub91} Yu. Kubyshin, "Higgs potentials from multidimensional 
Yang-Mills theories", {\em J. Math. Phys.}, {\bf 32}, (1991), 2200-2208.  

\bibitem{Kub} Yu. Kubyshin, "A Classification of Fibre Bundles 
over 2-dimensional Spaces", in: {\em New Developments in 
Algebraic Topology}, Proceedings of the Meeting, 
(Yu. Kubyshin et al., Eds.), University of the Algarve, 
Faro, 2000, p. 19-41. E-print archive: math.AT/9911217. 

\bibitem{NaSe} 
C. Nash, S. Sen, {\em Topology and Geometry for Physicists}, 
Academic Press, London, 1983.  

\bibitem{RP}
S. Randjbar-Daemi, R. Percacci, {\em Phys. Lett.}, {\bf 117B}, (1982), 
41.

\bibitem{Schw} A.S. Schwarz, {\em Commun. Math. Phys.}, {\bf 56}, 
(1977), 79-86. \\
V. Romanov, A. Schwarz, Yu. Tyupkin, {\em Nucl. Phys.}, {\bf B130}, 
(1977), 209-220. 

\bibitem{Span}
E.H. Spanier, {\em Algebraic Topology}, McGraw Hill, New York, 1966.

\bibitem{Steen}
N. Steenrod, {\em The Topology of Fibre Bundles}, Princeton University
Press, Princeton, 1951.

\bibitem{Swit}
R.M. Switzer, {\em Algebraic Topology - Homotopy and Homology}, 
Springer-Verlag, New York, 1975.

\bibitem{Tra} A. Trautman, {\em Rep. Math. Phys.}, {\bf 1}, (1970), 29. 

\bibitem{VK-book} I. Volobuev, Yu. Kubyshin, {\em Differential 
Geometry and Lie Algebras and Their Applications in Field Theory} 
(in Russian), Editorial URSS, Moscow, 1998.  

\bibitem{Wang} H.C. Wang, {\em Nagoya Math. J.}, {\bf 13}, (1958), 1.

\bibitem{Whi}
G.W. Whitehead, {\em Elements of Homotopy Theory},
Springer-Verlag, New York, 1978.

\bibitem{Wit77} E. Witten, {\em Phys. Rev. Lett.}, {\bf 38}, 
1977), 121-124.

\bibitem{Wit} 
E. Witten, {\em J. Geom. Phys.}, {\bf 9}, (1992), 303. 

\end{thebiblio}

\vspace{1cm}

\begin{itemize}
\item[1] {\small Departamento de Matem\'atica Aplicada IV. Universidad 
Polit\'ecnica de Catalu\~na. \\
M\'od. C-3, Campus Norte, c./ Jordi Girona 1-3, 
08034 Barcelona, Spain.}
\item[2] {\small Institute of Nuclear Physics, Moscow State University, 
119992 Moscow, Russia.}
\end{itemize}

\end{document}